\documentstyle[12pt]{article}

\renewcommand{\arraystretch}{1.0}

\renewcommand{\theequation}{\arabic{section}.\arabic{equation}}

\newcounter{leteq}
\newcommand{\steplet}{\stepcounter{leteq}\addtocounter{equation}{-1}}

\newenvironment{eqnalpha}{\setcounter{leteq}{1}
\renewcommand{\theequation}{\arabic{section}.\arabic{equation}\alph{leteq}}
\begin{eqnarray}}{\end{eqnarray}%
\renewcommand{\theequation}{\arabic{section}.\arabic{equation}}}

\newenvironment{eqnalphalabel}[1]{\setcounter{leteq}{1}
\raisebox{0cm}[0cm][0cm]{\begin{minipage}{1cm}%
\begin{eqnarray}\label{#1}&&\nonumber\end{eqnarray}\end{minipage}}
\renewcommand{\theequation}{\arabic{section}.\arabic{equation}\alph{leteq}}
\begin{eqnarray}}{\end{eqnarray}%
\renewcommand{\theequation}{\arabic{section}.\arabic{equation}}}

\newcommand{\spc}{{\ }}

\newcommand{\pp}{p}
\newcommand{\rr}{r}
\newcommand{\Pp}{P}
\newcommand{\RR}{R}
\newcommand{\rD}{_{\rm D}}

\begin{document}

\title   {
                  SDG fermion-pair algebraic SO(12) and Sp(10)   \\
                       models and their boson realizations.
         }
\author{
         {                  P. Navr\'atil\thanks
         { On leave of absence from the Institute of Nuclear Physics,
              Czech Academy of Sciences, \v{R}e\v{z} near Prague,
                               Czech Republic
         }                   and H. B. Geyer                   }\\
         {\it          Institute of Theoretical Physics,
                         University of Stellenbosch,           }\\
         {\it           Stellenbosch 7600, South Africa        }\\
         {                      J. Dobe\v{s}                   }\\
         {\it           Institute of Nuclear Physics,
                         Czech Academy of Sciences,            }\\
         {\it       CS 250 68 \v{R}e\v{z}, Czech Republic      }\\
         {                     J. Dobaczewski                  }\\
         {\it         Institute of Theoretical Physics,
                             Warsaw University,                }\\
         {\it        ul. Ho\.za 69, PL--00-681 Warsaw, Poland  }\\
         }

\maketitle

\begin{abstract}
It is shown how the boson mapping formalism may be applied as a
useful many-body tool to solve a fermion problem.  This is done
in the context of generalized Ginocchio models for which we
introduce S\mbox{-,} D\mbox{-,} and G\mbox{-pairs} of fermions
and subsequently construct the sdg-boson realizations of the
generalized Dyson type. The constructed SO(12) and Sp(10)
fermion models are solved beyond the explicit symmetry limits.
Phase transitions to rotational structures are obtained, also in
situations where there is no underlying SU(3) symmetry.
\end{abstract}



\newpage

\section{Introduction}
\label{sec1}
\setcounter{equation}{0}

It has often been recognized and stated that boson mapping is
not only relevant to discussions about the relationship between
phenomenological boson models and shell model type fermion
models, but that it also constitutes an attractive many-body
formalism in its own right (see e.g.{\spc}Refs.{\spc}\cite{DGH91,KM91} and
references therein).  The latter pronouncement is against the
background that one might profit from the use of boson variables
in the description of many-fermion problems through the
potentially simpler algebraic structures associated with bosons.
This philosophy has its roots in the very simple observation
that bosons commute to a $c$-number, while fermion pairs commute
to a non-trivial operator. Correspondingly the construction of
an orthonormal basis is in general then simpler for bosons than
for fermion pairs. Except for a few cases (see
e.g.{\spc}Refs.{\spc}[3--9]),
this has not been exploited much as a possible simpler route to
solution of a many-body problem.

In this paper we demonstrate how the known algebraic structure
associated with s\mbox{-,} d\mbox{-,} and g\mbox{-bosons} may
indeed be exploited to solve non-trivial fermion models.  We
analyze and discuss an extension of the original Ginocchio SO(8)
and Sp(6) models which have also been re-interpreted and used in
the Fermion Dynamical Symmetry Model (FDSM){\spc}\cite{FDSM}.  The
extension is based on a closed algebraic structure in terms of
fermion pairs which now also includes a G-pair and leads to the
SO(12) and Sp(10) dynamical symmetries.  Such an extension has
also recently been presented by P.~Feng {\it et al.}{\spc}\cite{FYZ91}
who discussed analytic expressions for energies in
some symmetry limits of the Sp(10) case.  Here we are able to
obtain solutions beyond the symmetry limits by using boson
mapping methods.  (See also Ref.{\spc}\cite{DNG93} for further
exploitation of the boson method in this context.)

Before we introduce and analyze these SDG-generalizations of the
Ginocchio (FDSM) models, we discuss in Sec.\ref{sec2} the basic
principle underlying the construction of algebraic fermion
models structured around the concept of favored pairs. We focus
namely on the respective roles of physical and algebraic angular
momenta{\spc}\cite{Dob87}, thus arriving at a division into active
and inert parts of the angular momentum from a point of view we
regard as more fundamental than the usual consideration where
such a division is simply introduced as a convenience.

In Sec.{\spc}\ref{sec3} we then construct generalizations of the
Ginocchio (FDSM) models for an arbitrary highest angular
momentum favored pair and subsequently discuss possible subgroup
chains and representations (Sec.{\spc}\ref{sec5}) as well as
realizations of these models in the nuclear shell-model space
(Sec.{\spc}\ref{sec4a}).  An equivalent description of
collective states in terms of the single-particle quadrupole
operator is discussed in Sec.{\spc}\ref{sec4b} where a
comparison with the harmonic oscillator quadrupole operator is
presented.

In Sec.{\spc}\ref{sec4} we construct the Dyson boson mapping
for the pair and multipole operators.  For the case of highest
angular momentum $J_{\rm{max}}$=4, namely for SDG-models, we
consider in Sec.{\spc}\ref{sec6} the role of possible spurious
states when the boson analysis is carried out in the complete
and unrestricted Fock space spanned by boson states of the
sdg-type.  The possible appearance and role of spurious states
when the full simplicity of a boson basis is exploited, is of
course well known and analyzed (see e.g.
Refs.{\spc}\cite{GEH86,DGH91,NG93}), but there are some detailed
properties for the present analysis worthwhile to be pointed out
and discussed.

The results of calculations we are able to carry out are
presented in Sec.{\spc}\ref{sec7}. In fact they can be relatively
easily obtained once the fermion problem is transformed to an
equivalent boson description.  We show that the SDG algebraic
models include phase transitions from anharmonic vibrations to
rotations.  In particular, we obtain well-pronounced rotational
solutions in the SO(12) model, which are not related to any
SU(3) symmetry limit.  In Sec.{\spc}\ref{sec8} we give a summary
and formulate some conclusions.

\section{Collective algebraic models and angular momentum}
\label{sec2}
\setcounter{equation}{0}

In algebraic collective models one assumes that collective
states (of a given collective mode) can be constructed by acting
with the generators $A$ of a so-called spectrum generating
algebra{\spc}\cite{SGA88} ${\cal A}$ on a reference state $|{\rm
ref}\rangle$.  The reference state is a fermion state (not
necessarily the ground state) and ${\cal A}$ is an algebra of
fermion operators.

Such models have the advantage of being simple enough to
incorporate in them an exact conservation of basic symmetries of
fermion systems. The simplest way to do that is to build the
generators of ${\cal A}$ as irreducible tensors of conserved
symmetries. In particular, the generators should either conserve
or change by a given value the particle number of the system,
and similarly, either conserve or carry with them a given
angular momentum.  In what follows we concentrate our discussion on
the conservation of these two symmetries.

The conservation of the particle number means that some subset
of generators of ${\cal A}$ should commute with the particle
number operator and constitute a spectrum generating algebra for
a given fermion system. Some other generators should decrease or
increase the fermion number either by one, in which case one
unifies the dynamics of even and odd nuclei, or by two if one
wants to consider only even systems.  In this paper we consider
only spectrum generating algebras built of pair-creation,
pair-annihilation and single-particle operators. In principle
one should consider many-body terms in either of these operators
which would be a natural extension in the spirit of constructing
effective operators resulting from a reduction to a small phase
space. In the algebraic approach, however, this is a very
difficult task and only a few developments have been made in
this direction (see e.g. Ref.{\spc}\cite{Dob90}).

Let us attribute to every generator $A_{NJM}$ of ${\cal A}$ the
angular momentum quantum number $J$ and its projection $M$, and
let $N$ enumerate generators with the same $J$ and $M$, if
necessary.  The assumption that the generators $A_{NJM}$ are
irreducible rotational tensors can be formulated in terms of the
following commutation relation
   \begin{equation}\label{e1}
   \big[J_{1M},A_{NJM'}\big] = -\sqrt{J(J+1)}(JM1M'|JM+M')A_{NJM+M'},
   \end{equation}
where $J_{1M}$ are spherical components of the angular momentum
operator{\spc}\cite{VMK88}. We will call $J_{1M}$ the physical angular
momentum to distinguish it from other angular momenta which
appear later on.  The physical angular momentum is a standard
operator acting on the coordinates and spins of all particles in
the fermion system considered, and it obeys standard so(3)
commutation relations\footnote
{The sign of the right-hand side of Eq.{\spc}(\ref{e3})
corresponds to the standard choice{\spc}\cite{VMK88} for relative
phases of components of the angular momentum operator $J_{1M}$.
Similarly, the sign in Eq.{\spc}(\ref{e1}) corresponds to a
certain assumed phase convention for the $A_{NJM}$ generators.},
   \begin{equation}\label{e3}
   \big[J_{1M},J_{1M'}\big] = -\sqrt{2}(1M1M'|1M+M')J_{1M+M'} .
   \end{equation}
We denote this algebra by so$_J$(3) to indicate that it is
generated by the operators $J_{1M}$.

The commutation relation (\ref{e1}) is easy to implement
whenever the generators $A_{NJM}$ are given in terms of the
space coordinates and spins. Equivalently, one can construct the
generators $A_{NJM}$ from fermion creation and annihilation
operators by coupling them to a given angular momentum. In any
case, a commutation relation such as (\ref{e1}) assures the
correct transformation properties of generators under rotations
of the reference frame.  It also allows one to determine the
angular momentum of a collective state by coupling together
angular momenta of the generators $A_{NJM}$.

Let us assume that in the algebra ${\cal A}$ there exists a
$J$=1 operator $A_{11M}$$\equiv$${\Pp}_{1M}$ which acts on the
labels of generators as an angular momentum{\spc}\cite{Dob87}, i.e.,
   \begin{equation}\label{e2}
   \big[{\Pp}_{1M},A_{NJM'}\big] =
      -\sqrt{J(J+1)}(JM1M'|JM+M')A_{NJM+M'} .
   \end{equation}
Applying this relation to $A_{11M}$ we see that ${\Pp}_{1M}$
also fulfills the so(3) commutation relations as in
Eq.{\spc}(\ref{e3}).  We will call ${\Pp}_{1M}$ the algebraic
angular momentum and denote the corresponding algebra by
so$_{\Pp}$(3).

Assumption (\ref{e2}) is the key element for constructing
collective fermion-pair algebras.  Even if it constitutes a
rather stringent condition imposed upon the algebra ${\cal A}$,
it is a very natural one if one aims at constructing a useful
algebraic model.  It means that we want to include in ${\cal A}$
the operator which would measure the physical angular momentum.
In this way the model space, which is generated by acting with
generators of ${\cal A}$ on the reference state, will always
comprise complete multiplets of the physical angular momentum.
By the same requirement we also have to assume that ${\Pp}_{1M}$
is a single-particle operator, similarly to $J_{1M}$.

Of course, the physical angular momentum $J_{1M}$ itself can
belong to the spectrum generating algebra ${\cal A}$.  Eqs.
(\ref{e1}) and (\ref{e2}) are then just a duplication of the
same property.  This is the case, for example, in the CM(3)
model{\spc}\cite{WBC73} or the symplectic Sp(3,R) model{\spc}\cite{RR77}
where generators are constructed from particle coordinates.
However, in what follows we need not distinguish between the two
separate cases when $J_{1M}$ and ${\Pp}_{1M}$ are equal or not,
because the former can be considered as a special case of the
latter -- see also below.  When $J_{1M}$ and ${\Pp}_{1M}$ are
different operators we have to consider a larger algebra,
so$_J$(3)$+{\cal A}$, comprising both the physical spin $J_{1M}$
and the spectrum generating generators $A_{NJM}$.  It is,
however, useful to arrange generators of this algebra in another
way.

Subtracting Eqs.{\spc}(\ref{e1}) and (\ref{e2}) we see that all
generators $A_{NJM}$ commute with the angular momentum operator
${\RR}_{1M}$ which is the difference of the physical and
algebraic angular momentum:
   \begin{equation}\label{e4}
   {\RR}_{1M} := J_{1M} - {\Pp}_{1M} ,
   \end{equation}
   \begin{equation}\label{e6}
   \big[{\RR}_{1M'},A_{NJM}\big] = 0 .
   \end{equation}
Therefore, we obtain a direct sum ${\cal A'}$,
   \begin{equation}\label{e5}
   {\cal A'} = {\mbox{\rm so$_{\RR}$(3)}}\oplus{\cal A} ,
   \end{equation}
of the so$_{\RR}$(3) algebra composed of the operators
${\RR}_{1M}$, and of the spectrum generating algebra ${\cal A}$.
We will call ${\RR}_{1M}$ the inactive or inert angular momentum.

Moreover, applying Eq.{\spc}(\ref{e6}) to
${\Pp}_{1M}$$\equiv$$A_{11M}$ we see that
   \begin{equation}\label{e11}
   \big[{\RR}_{1M'},{\Pp}_{1M}\big] = 0 ,
   \end{equation}
i.e., in the algebra ${\cal A'}$ there exist {\it two} commuting
angular momentum operators.  This fact has important
consequences for the structure of the underlying fermion space.
Indeed, having at our disposal two commuting single-particle
operators we can attribute {\it two} angular momentum quantum
numbers to every single-particle state.  More specifically, let
us suppose that algebra ${\cal A}$ is built in the spherical
shell model space of single-fermion states described by the
creation operators $a^+_{njm_j}$.  These states are grouped in
multiplets ($j$-shells) with respect to the physical angular
momentum and therefore $a^+_{njm_j}$ are irreducible tensors
with respect to $J_{1M}$,
   \begin{equation}\label{e12}
   \big[J_{1M},a^+_{njm_j}\big] =
      -\sqrt{j(j+1)}(1Mjm_j|jm_j+M)a^+_{njm_j+M},
   \end{equation}
where $n$ distinguishes multiplets of the same $j$.  The
existence of two commuting angular momenta allows one now to
group single-particle states in another set of multiplets
(${\rr}$-${\pp}$-shells) each described the by fermion
creation operators $a^+_{n'{\rr}m_{\rr}{\pp}m_{\pp}}$ which are
irreducible tensors with respect to ${\RR}_{1M}$ and
${\Pp}_{1M}$ {\it simultaneously},
   \begin{eqnalpha}
   \big[{\RR}_{1M},a^+_{n'{\rr}m_{\rr}{\pp}m_{\pp}}\big] &=&
      -\sqrt{{\rr}({\rr}+1)}(1M{\rr}m_{\rr}|{\rr}m_{\rr}+M)
                    a^+_{n'{\rr}m_{\rr}+M{\pp}m_{\pp}}, \label{e15a} \\
                                                        \steplet
   \big[{\Pp}_{1M},a^+_{n'{\rr}m_{\rr}{\pp}m_{\pp}}\big] &=&
      -\sqrt{{\pp}({\pp}+1)}(1M{\pp}m_{\pp}|{\pp}m_{\pp}+M)
                    a^+_{n'{\rr}m_{\rr}{\pp}m_{\pp}+M}. \label{e15b}
   \end{eqnalpha}%
Here $n'$ distinguishes the ${\rr}$-${\pp}$-shells with the same
values of ${\rr}$ and ${\pp}$. Since the physical angular momentum
is by definition the sum of the active and inactive ones,
   \begin{equation}\label{e13}
   J_{1M} = {\Pp}_{1M} + {\RR}_{1M},
   \end{equation}
(cf.  Eq.{\spc}(\ref{e4})), the $j$-shells and
the ${\rr}$-${\pp}$-shells are related by the angular momentum
coupling,
   \begin{eqnalpha}
   a^+_{njm} &=& \sum_{m_{\rr}m_{\pp}} ({\rr}m_{\rr}{\pp}m_{\pp}|jm)
               a^+_{n'{\rr}m_{\rr}{\pp}m_{\pp}} , \label{e14a} \\
                                                        \steplet
   a^+_{n'{\rr}m_{\rr}{\pp}m_{\pp}} &=&
                 \sum_{jm_j} ({\rr}m_{\rr}{\pp}m_{\pp}|jm)
                 a^+_{njm} , \label{e14b}
   \end{eqnalpha}%
and the quantum number $n$ comprises three quantum numbers
$n'$, ${\rr}$ and ${\pp}$.

We have therefore shown that assumption (\ref{e2}) induces a
very specific structure of the fermion space, namely, the
grouping of single-particle states into ${\rr}$-${\pp}$-shells.
The values of the quantum numbers ${\rr}$ and ${\pp}$ are as yet
unspecified, but one can easily see that by using only the
${\rr}$=0 value one recovers the original structure of
$j$-shells because one then has ${\pp}$=$j$ and $n'$=$n$.  In
this case all fermion creation and annihilation operators
commute with the inactive angular momentum, cf.
Eq.{\spc}(\ref{e15a}).  This situation is therefore identical to
the one where the physical angular momentum $J_{1M}$ itself
belongs to the algebra ${\cal A}$ and the inactive angular
momentum ${\RR}_{1M}$ {\spc}(\ref{e4}) vanishes.

Every ${\rr}$-${\pp}$-shell contains states with physical
angular momenta $j$=$|{\rr}$$-$${\pp}|,\ldots,{\rr}$+${\pp}$.
The structure of nuclear shells can therefore be embedded in the
${\rr}$-${\pp}$ scheme in several ways.  For example, the
$N_0$-th major oscillator shell contains states with
$j$=$\frac{1}{2}$, $\frac{3}{2},\ldots,\frac{2N_0+1}{2}$ and can
be identified with the single ${\rr}$-${\pp}$-shell either for
${\pp}$=$\frac{N_0+1}{2}$ and ${\rr}$=$\frac{N_0}{2}$ or for
${\pp}$=$\frac{N_0}{2}$ and ${\rr}$=$\frac{N_0+1}{2}$ The same
nuclear shell can also be identified with two or more
${\rr}$-${\pp}$-shells.  For example, if one wants to take
into account the fact that the highest $j$=$\frac{2N_0+1}{2}$
multiplet of the $N_0$-th oscillator shell (the intruder state) is
well separated from the other ones, one may consider it to be an
${\rr}$-scalar shell and identify the remaining multiplets as
another ${\rr}$-${\pp}$-shell.

The fact that the fermion space is composed of
${\rr}$-${\pp}$-shells allows one to construct some useful
collective spectrum generating algebras ${\cal A}$.  We may
start by constructing a large algebra ${\cal A''}$ in the
fermion space of the ${\rr}$-${\pp}$-shells and then select only
those generators which are scalar with respect to the inactive
angular momentum ${\RR}_{1M}$.  In this way we fulfill by
construction assumption (\ref{e2}).  In the next Section we
present such a derivation for the case of a fermion space
composed of a single ${\rr}$-${\pp}$-shell.  A generalization to
the case of several ${\rr}$-${\pp}$-shells is straightforward
and will not be discussed here.

The construction in the next Section parallels the original
derivation by Ginocchio{\spc}\cite{Gin80} who follows the ideas
introduced in the pseudo-spin model{\spc}\cite{HA69} and uses the
coupling of the {\it half-integer} pseudo-spin $i$ with the {\it
integer} pseudo-orbital angular momentum $k$ to the total
angular momentum $j$.  His starting point is the fermion space
composed of a single $k$-$i$-shell in which he considers either
$k$ or $i$ to be the inactive angular momentum.  The cases
$i$=$\frac{3}{2}$ for inactive $k$ and $k$=1 for inactive $i$
are then discussed in detail. In what follows we use the
notation focused on inactive and active angular momentum ${\rr}$
and ${\pp}$, respectively, rather then on integer and
half-integer angular momenta $k$ and $i$, as the former
classification seems to be more fundamental.

\section {Generalization of the Ginocchio model}
\label{sec3}
\setcounter{equation}{0}

In this Section we generalize the Ginocchio prescription{\spc}\cite{Gin80}
to construct fermion-pair algebras in such a way
that they incorporate fermion pairs up to an arbitrary even
angular momentum $J_{\rm{max}}$.  We discuss the application
to the case of $J_{\rm{max}}$=4, i.e., $S$, $D$, and $G$
pairs, in some more detail (see also Ref.{\spc}\cite{FYZ91}).

Consider the fermion Fock space built on a single
${\rr}$-${\pp}$-shell described by the fermion creation operators
$a^+_{{\rr}m_{\rr}{\pp}m_{\pp}}$, where ${\rr}$ and
${\pp}$ are inactive and active angular momenta, respectively.
By the vector coupling of the ${\rr}m_{\rr}$ and
${\pp}m_{\pp}$ indices one obtains the inactive and active
angular momentum operators in the explicit form
   \begin{eqnalpha}
   {\RR}_{1M} &:=& -\hat{{\rr}}\hat{{\pp}}\sqrt{{\rr}({\rr}+1)/3}
               {\displaystyle\sum_{m_{\rr}m'_{\rr}m_{\pp}m'_{\pp}}}\!\!\!
               ({\rr}m_{\rr}{\rr}m'_{\rr}|1M)({\pp}m_{\pp}{\pp}m'_{\pp}|00)
               a^+_{{\rr}m_{\rr}{\pp}m_{\pp}}
               \tilde{a}_{{\rr}m'_{\rr}{\pp}m'_{\pp}} , \label{e7b}  \\
                                                        \steplet
   {\Pp}_{1M} &:=& -\hat{{\rr}}\hat{{\pp}}\sqrt{{\pp}({\pp}+1)/3}
               {\displaystyle\sum_{m_{\rr}m'_{\rr}m_{\pp}m'_{\pp}}}\!\!\!
               ({\rr}m_{\rr}{\rr}m'_{\rr}|00)({\pp}m_{\pp}{\pp}m'_{\pp}|1M)
               a^+_{{\rr}m_{\rr}{\pp}m_{\pp}}
                 \tilde{a}_{{\rr}m'_{\rr}{\pp}m'_{\pp}} , \label{e7a}
   \end{eqnalpha}%
where
   \begin{equation}\label{e9}
   \tilde{a}_{{\rr}m_{\rr}{\pp}m_{\pp}} :=
      (-1)^{{\rr}-m_{\rr}}(-1)^{{\pp}-m_{\pp}}
         a_{{\rr},-m_{\rr}{\pp},-m_{\pp}} ,
   \end{equation}
and the hat over any angular momentum quantum number is
understood as $\hat{{\rr}}$=$\sqrt{2{\rr}+1}$.  The summation
over $m_{\rr}$ and $m'_{\rr}$ in (\ref{e7a}), and over $m_{\pp}$
and $m'_{\pp}$ in (\ref{e7b}) couples to total spins ${\RR}$=0
and ${\Pp}$=0, respectively. ${\Pp}_{1M}$ is therefore scalar
with respect to the angular momentum ${\RR}_{1M}$ and vice
versa, i.e., Eq.{\spc}(\ref{e11}) is fulfilled.

The bifermion creation operators
$a^+_{{\rr}m_{\rr}{\pp}m_{\pp}}a^+_{{\rr}m'_{\rr}{\pp}m'_{\pp}}$,
together with their hermitian conjugates
$a_{{\rr}m'_{\rr}{\pp}m'_{\pp}}a_{{\rr}m_{\rr}{\pp}m_{\pp}}$ and
all commutators of the two sets, give the standard bifermion
so$(2(2{\rr}+1)(2{\pp}+1))$ algebra.  The spectrum generating
algebra ${\cal A}$ can be obtained as a subalgebra of
so$(2(2{\rr}+1)(2{\pp}+1))$ composed of combinations of
generators which are scalars with respect to the inactive
angular momentum ${\RR}_{1M}$. Therefore, ${\cal A}$ is given by
the bifermion creation operators,
   \begin{equation}\label{e10}
   F^+_{JM} =\sqrt{\frac{g}{2}}
         {\displaystyle\sum_{m_{\rr}m'_{\rr}m_{\pp}m'_{\pp}}}
             ({\rr}m_{\rr}{\rr}m'_{\rr}|00)
             ({\pp}m_{\pp}{\pp}m'_{\pp}|JM)
                  a^+_{{\rr}m_{\rr}{\pp}m_{\pp}}
                  a^+_{{\rr}m'_{\rr}{\pp}m'_{\pp}} ,
   \end{equation}
their hermitian conjugates $F_{JM}$, and all their commutators.
In expression (\ref{e10}) we have included an arbitrary
normalization factor $g$ in terms of which the two-particle
states are normalized as (cf.{\spc}Ref.{\spc}\cite{DGH91})
   \begin{equation}\label{e23}
   \langle0|F_{JM}F^+_{J'M'}|0\rangle = g\delta_{JJ'}\delta_{MM'} .
   \end{equation}

Since the physical single-particle (half-integer) angular
momentum $j$ is the sum of the inactive angular momentum ${\rr}$
and of the algebraic angular momentum ${\pp}$, we only consider
here cases when either ${\rr}$ is integer and ${\pp}$
half-integer, or vice versa.

As shown in the Appendix, the algebraic structure of pair
operators $F^+_{JM}$ is equivalent to that of {\em fermion}
pairs occupying a single-${\pp}$ shell for half-integer ${\pp}$,
or to that of {\em boson} pairs occupying a single-${\pp}$ shell
for integer ${\pp}$.  The former gives the orthogonal algebra
so(4${\pp}$+2) while the latter gives the symplectic algebra
sp(4${\pp}$+2).  The single-${\pp}$ shells are repeated
2${\rr}$+1 or (2${\rr}$+1)/2 times, respectively, and the copies
are numbered by the magnetic quantum number $m_{\rr}$ of the
inactive angular momentum ${\rr}$.

In a single-${\pp}$ shell for either half-integer or integer
${\pp}$, bosons or pairs of fermions can only be coupled to even
total spins $J$. Therefore, the spectrum generating algebra
${\cal A}$ is in both cases comprised of pairs of fermions
$F^+_{JM}$ for $J$=0,2,$\ldots,$$J_{\rm{max}}$, where
$J_{\rm{max}}$=2$[{\pp}]$ (doubled integer part of ${\pp}$).

Algebraic properties of the models constructed here are,
therefore, the same as those of particles restricted to a
single-${\pp}$ shell.  On the other hand, the Hamiltonian is a
scalar with respect to the physical angular momentum and couples
different copies of the single-${\pp}$ shells, giving rise to
rather rich dynamic properties.

Commutation relations of ${\cal A}$ can be obtained by either
considering the underlying single-${\pp}$ structure or by an
explicit calculation. The latter method can be facilitated by
using useful techniques developed in Ref.{\spc}\cite{CCK93}.  One
then obtains the so(4${\pp}$+2) and sp(4${\pp}$+2) commutation
relations for half-integer and integer ${\pp}$, respectively, in
the form:
   \begin{equation}
   \left[F^+_{J_2M_2},\left[F_{J_1M_1},F^+_{J_2'M_2'}\right]\right] =
   \sum_{J_1'M_1'} C^{J_2M_2J_2'M_2'}_{J_1M_1J_1'M_1'}F^+_{J_1'M_1'} ,
   \end{equation}
where the structure constants read:
   \begin{eqnarray}\label{e26}
   C^{J_2M_2J_2'M_2'}_{J_1M_1J_1'M_1'} &=& \displaystyle{
      \frac{4g}{{\hat {\rr}}^2}\hat{J_1}\hat{J_2}\hat{J_1'}\hat{J_2'}
        \sum_{J''=0,1,\ldots,2{\pp}} (J_1M_1J_1'M_1'|J''M'')
                                   (J_2M_2J_2'M_2'|J''M'')}
   \left\{\begin{array}{ccc} {\pp}    & {\pp}    & J_1  \\
                             {\pp}    & {\pp}    & J_1' \\
                             J_2  & J_2' & J''       \end{array}\right\}
                                    \nonumber \\
                                    &=& \displaystyle{
   \frac{4g}{{\hat {\rr}}^2}
        \sum_{m_{\pp}n_{\pp}m_{\pp}'n_{\pp}'}
           ({\pp}m_{\pp}{\pp}n_{\pp} |J_1M_1)
           ({\pp}m_{\pp}'{\pp}n_{\pp}'|J_1'M_1')
                  ({\pp}m_{\pp}'{\pp}m_{\pp}|J_2M_2)
                  ({\pp}n_{\pp}'{\pp}n_{\pp}|J_2'M_2')} .
                                    \nonumber \\ &&
   \end{eqnarray}

These same commutation relations can be presented in a different
form by explicitly introducing the one-body operators of ${\cal A}$,
   \begin{equation}\label{e22}
   P_{JM} =  h{\displaystyle\sum_{m_{\rr}m'_{\rr}m_{\pp}m'_{\pp}}}
             ({\rr}m_{\rr}{\rr}m'_{\rr}|00)({\pp}m_{\pp}{\pp}m'_{\pp}|JM)
                    a^+_{{\rr}m_{\rr}{\pp}m_{\pp}}
             \tilde{a}_{{\rr}m'_{\rr}{\pp}m'_{\pp}} ,
   \end{equation}
which can have angular momenta $J$=0,1,2,$\ldots,$$2{\pp}$.
Again, we include in the definition of $P_{JM}$ an arbitrary
normalization factor $h$, which can facilitate the comparison of
different conventions used in the literature.  For example, the
factor which gives the operator $P_{1M}$ normalized as an
angular momentum, Eq.{\spc}(\ref{e7a}), is
$h$=$-\hat{{\rr}}\hat{{\pp}}\sqrt{{\pp}({\pp}+1)/3}$.  The
so(4${\pp}$+2) and sp(4${\pp}$+2) commutation relations now read
   \begin{eqnalphalabel}{eq2.2}
   \left[ F_{JM}, F^+_{J'M'}\right] & = &
   g\delta_{JJ'}\delta_{MM'} + (-1)^{2{\pp}}\frac{2g}{h}
      \frac{\hat{J}\hat{J'}}{\hat{{\rr}}}(-1)^{J+M} \nonumber \\
   &\times& \sum_{J''=0,1,\ldots,2{\pp}}(J,-MJ'M'|J''M'')
   \left\{ \begin{array}{ccc} J & J' & J''\\{\pp}&{\pp}&{\pp}
             \end{array}\right\}
                                               P_{J''M''} \\
                                                        \steplet
   \left[ P_{JM}, F^+_{J'M'}\right] & = &
      -2h(-1)^{2{\pp}}\frac{\hat{J}\hat{J'}}{\hat{{\rr}}} \nonumber \\
   &\times& \sum_{J''=0,2,\ldots,2[{\pp}]} (JMJ'M'|J''M'')
      \left\{ \begin{array}{ccc} J & J' & J''\\{\pp}&{\pp}&{\pp}
                \end{array}\right\}
                                              F^+_{J''M''} \\
                                                        \steplet
     \left[ P_{JM}, P_{J'M'}\right] & = &
   h(-1)^{2{\pp}}\frac{\hat{J}\hat{J'}}{\hat{{\rr}}}
                                                 \nonumber \\
   &\times& \sum_{J''=0,1,\ldots,2{\pp}}
              ((-1)^{J+J'}-(-1)^{J''}) (JMJ'M'|J''M'')
   \left\{ \begin{array}{ccc} J & J' & J''\\{\pp}&{\pp}&{\pp}
             \end{array}\right\}
                                             P_{J''M''}
                                                 \nonumber \\
   \end{eqnalphalabel}%

We conclude this section by presenting the generators ${\cal A}$
in terms of physical angular momenta of single particle states,
i.e., in terms of creation and annihilation operators of the
$j$-shells, Eq.{\spc}({\ref{e15a}}).  Recoupling the angular
momenta one obtains the pair operators (\ref{e10}) and the
one-body operators (\ref{e22}) in the form:
   \begin{eqnalpha}
   F^+_{JM} &=&
   \sqrt{\frac{g}{2}} \sum_{j_1j_2}(-1)^{J+{\pp}+{\rr}+j_1}
   \frac{\hat{j_1}\hat{j_2}}{\hat{{\rr}}}
   \left\{\begin{array}{ccc} j_1 & j_2 & J \\
                             {\pp}   & {\pp}   & {\rr}    \end{array}\right\}
   \left(a^+_{j_1}a^+_{j_2}\right)^{(J)}_{M} , \label{e86a} \\
                                                        \steplet
   P_{JM} &=&
   h \sum_{j_1j_2}(-1)^{J+{\pp}+{\rr}+j_1}
   \frac{\hat{j_1}\hat{j_2}}{\hat{{\rr}}}
   \left\{\begin{array}{ccc} j_1 & j_2 & J \\
                             {\pp}   & {\pp}   & {\rr}    \end{array}\right\}
   \left(a^+_{j_1}\tilde{a}_{j_2}\right)^{(J)}_{M} . \label{e86b}
   \end{eqnalpha}%

\section{Subgroup chains and representations in the model space}
\label{sec5}
\setcounter{equation}{0}

{}From the commutation relations (\ref{eq2.2}) one can deduce
the subgroup chain structure of the generalized Ginocchio
(FDSM) model. For the case $K=0$, $i\equiv p$ one has
\begin{equation}\label{eq4.1}
{\rm SO}(2(2i+1))\supset {\rm U}(2i+1)\supset
{\rm Sp}(2i+1)\supset {\rm SO}(3)
\end{equation}
and
\begin{equation}\label{eq4.2}
{\rm SO}(2(2i+1))\supset {\rm Sp}(2i+1)\otimes {\rm SU}(2)\supset
{\rm SO}(3)\otimes {\rm SU}(2)\; ,
\end{equation}
while for the $I=0$, $p\equiv k$ case the subgroup chain
structure is
\begin{equation}\label{eq4.3}
{\rm Sp}(2(2k+1))\supset {\rm U}(2k+1)\supset
{\rm SO}(2k+1)\supset {\rm SO}(3)
\end{equation}
and
\begin{equation}\label{eq4.4}
{\rm Sp}(2(2k+1))\supset {\rm SO}(2k+1)\otimes {\rm SU}(2)\supset
{\rm SO}(3)\otimes {\rm SU}(2)
\end{equation}

The groups U($2p+1$) in relations (\ref{eq4.1}) and
(\ref{eq4.3}) are generated by the multipole operators $P_{JM}$
in (\ref{e22}). The generators of the groups Sp($2p+1$) and
SO($2p+1$) are the multipole operators $P_{JM}$ with $J$ odd and
$p\equiv i$ in the former and $p\equiv k$ in the latter case.
The angular momentum group SO(3) is generated by $P_{1M}$ and
the seniority group SU(2) in the chains (\ref{eq4.2}) and
(\ref{eq4.4}) by $F^+_{00}$, $F_{00}$ and $P_{00}$.

For the SO(8) model another algebraic chain is possible, namely
one which includes SO(7). This possibility is linked to the
accidental zero of a 6j-symbol,
\begin{equation}\label{eq4.5}
\left\{ \begin{array}{ccc} 2 & 2 & 2\\3/2\;&3/2\;&3/2 \end{array}\right\}
=0
\end{equation}
and the corresponding simplification of the relations
(\ref{eq2.2}).

We now turn to the SO(12) and Sp(10) models in some more detail.
For these models the fermion model space is obtained by
successive action of the respective $S$, $D$, and $G$
pair creation operators on the fermion vacuum.  These states
could also be obtained by combining configurations of the active
and inert spaces. Only those states of the inert space are
considered in which single-particle pseudo-orbital angular
momenta or pseudospins are pairwise coupled to 0. For the $k=2$
and $i=\frac{5}{2}$ active parts, the basic building blocks are
thus characterized by the symmetries [11] and [2] in the
respective inert spaces.  The total wave function should of
course be antisymmetric and if states are classified according
to Young tables, conjugate Young schemes in the active and inert
spaces are combined. As the wave function in the inert space has
either $I$ or $K=0$, while in the decomposition of the product
of two conjugate Young schemes the totally antisymmetric
representation occurs just once, states in the fermion model
space are uniquely denoted by the allowed labels corresponding
to the active space.

For the Sp(10) case, the $k=2$ active part lies in the SU(5)
space whereas the inert part is in the SU($\frac{2\Omega}{5}$)
space.  The possible representations for the inert part are
given by the Young labels $[f_1,f_2,\ldots,f_{2\Omega/5}]$ with
$5\geq f_1=f_2\geq f_3=f_4\geq\ldots\geq f_{2\Omega/5-1}=
f_{2\Omega/5}\geq 0$ and $f_1+f_2+\ldots+f_{2\Omega/5} =2N$, $N$
denoting the number of fermion pairs. Allowed conjugate
representations in the active SU(5) space have the Young labels
$[2N-2p-4q-6r-8s,2p+2q+2r+2s,2q+2r+2s,2r+2s,2s]\equiv
[2N-2p-4q-6r-10s,2p+2q+2r,2q+2r,2r]$. Here, $p,q,r,$ and $s$ are
non-negative integers such that $4p+6q+8r+10s\leq 2N$. Moreover,
the dimensionality of the inert space restricts  SU(5)
representations in the active space to $2N-2p-4q-6r-8s\leq
\frac{2\Omega}{5}$. This is just the Pauli principle
restriction.

Analogous considerations hold for the SO(12) case with the
$i=\frac{5}{2}$ active part in the SU(6) space.  The possible
representations for the SU($\frac{2\Omega}{6}$) inert part have
the Young labels $[2f_1,2f_2,\ldots,2f_{2\Omega/6}]$ with $6\geq
2f_1\geq 2f_2\geq\ldots\geq 2f_{2\Omega/6}\geq 0$ and
$2f_1+2f_2+\ldots+2f_{2\Omega/6}=2N$.  Conjugate representations in
the active SU(6) space are denoted by the Young labels
$[N-j-2k,N-j-2k,j+k,j+k,k,k]\equiv [N-j-3k,N-j-3k,j,j,0]$. Here,
$j$ and $k$ are non-negative integers such that $2j+3k\leq N$.
The dimensionality of the inert space now restricts SU(6)
representations in the active space to $N-j-2k\leq
\frac{2\Omega}{6}$.

\section{Links to the nuclear shell-model space}
\label{sec4a}
\setcounter{equation}{0}

The major nuclear shells consist of normal-parity subshells,
with values of single-particle angular momentum ranging from
$j$=$\frac{1}{2}$ to $j$=$\frac{2N_0-1}{2}$ where $N_0$ is the
radial quantum number, and of the unique-parity orbitals with
$j$=$\frac{2N_0+3}{2}$.  In Table \ref{table1} we give for $N_0$=3,
4, 5, and 6  possible identifications of the normal-parity
subshells with the $\rr$-$\pp$-shells.  All cases involving
fermion pairs up to $J_{\rm{max}}$=6 are enumerated. In
addition, we also list identifications which leave apart the
smallest subshell of $j$=$\frac{1}{2}$. This would require
treating this subshell as an $\rr$-scalar.  In all cases the
unique-parity orbitals have to be treated separately and in the
algebraic models are always considered to be the $\rr$-scalars.
Identifications exploited in the FDSM{\spc}\cite{FDSM}
are denoted by asterisks.

One can see that the Sp(10) SDG fermion-pair algebraic models
can be constructed for $N_0$=4 and 5, and the SO(12) SDG models
for $N_0$=5 and 6. It is interesting to note that both models can
be constructed in the same $N_0$=5 shell-model space.  If one
leaves out the $j$=$\frac{1}{2}$ shell, the SO(12) model can
also be constructed for $N_0$=4 and the Sp(10) model can be
constructed for $N_0$=6.

The SD SO(8) and Sp(6) models of the FDSM can therefore be
generalized to incorporate the G fermion pairs. In particular,
in the $N_0$=4, 5, and 6 major shells all four models, SO(8),
Sp(6), SO(12), and Sp(10), can be constructed and compared,
provided that in some cases one accepts to leave apart the
smallest subshell.  In Sec.{\spc}\ref{sec7} we discuss the $N_0$=6
case in some more detail.

\section{Quadrupole single-particle operator}
\label{sec4b}
\setcounter{equation}{0}

In the algebraic collective models discussed all possible
$N$-pair states can be obtained by acting on the fermion vacuum
with the pair creation operators,
   \begin{equation}\label{e82}
   |\Psi\rangle = F^+_{J_1M_1}\ldots F^+_{J_NM_N}|0\rangle .
   \end{equation}
Therefore, one often considers these operators as basic building
blocks which allow for a complete description of the collective
space. On the other hand, as proposed by the Single-particle
Coherent Excitation Model (SCEM){\spc}\cite{Dob90a,DR90}, one can
also consider the quadrupole single-particle operator
$Q_M$$\equiv$$P_{2M}$ and the monopole pair
$S^+$$\equiv$$F^+_{00}$ as another set of building blocks.
Then any other pair creation operator $F^+_{JM}$ can
be expressed as a multiple commutator of $Q_M$ with $S^+$ (see
Eq.{\spc}(\ref{eq2.2}b)),
   \begin{equation}\label{e84}
   F^+_{JM} \sim \left([Q\cdot[Q\cdot\ldots[Q\cdot,S^+]\ldots]]\right)_{JM},
   \end{equation}
with an appropriate angular momentum coupling. In particular,
the stretched coupling scheme can be used to raise the angular
momentum of fermion pairs:
   \begin{equation}\label{e90}
    \left([Q,F^+_J]\right)^{(J+2)}_{M+M'} =
      -2\sqrt{5}h(-1)^{2p}\frac{\hat{J}}{\hat{r}}
       \left\{\begin{array}{ccc} 2 & J & J+2 \\
                                 p & p & p     \end{array}\right\}
       F^+_{J+2,M+M'} .
   \end{equation}
In this way, all pairs for $J$$>$0 can be obtained because the
6-$j$ symbol involved here does not vanish{\spc}\cite{VMK88} as long
as $J$$<$2$[{\pp}]$. The complete collective algebra can
therefore be recovered from the operators $Q$ and $S^+$.

As a direct consequence of this fact we observe that all
collective states can be obtained as
   \begin{equation}\label{e83}
   |\Psi\rangle = Q_{M_1}\ldots Q_{M_{K_1}}S^+\ldots
                  Q_{M_1}\ldots Q_{M_{K_N}}S^+|0\rangle ,
   \end{equation}
i.e., as a sequence of $N$ monopole-pair creation operators
$S^+$ interspersed with quadrupole operators $Q_M$, and the sets
of states (\ref{e82}) and (\ref{e83}) are identical. The
total number of quadrupole operators in the sequence can be
large, and is only limited by the angular momentum accessible in
the collective space.

Similarly, all generators $P_{JM}$ constituting the core
subalgebra can be obtained from multiple commutators of $Q_M$,
see Eq.{\spc}(\ref{eq2.2}c).  Hence, in cases when the
$S^+$\mbox{-pair} condensate contains components in all
irreducible representations of the core subalgebra, all
collective states can also be obtained from
   \begin{equation}\label{e91}
   |\Psi\rangle = Q_{M_1}\ldots Q_{M_{K}}(S^+)^N|0\rangle .
   \end{equation}

For the SD algebraic models we have at our disposal the set of
six building blocks $S^+$ and $D^+_M$=$F^+_{2M}$ and another
equivalent set of six building blocks $S^+$ and $Q_M$.  However,
for models which use more pairs, like the SDG SO(12) and Sp(10)
models discussed in the present paper, the number of pair
creation operators $F^+_{JM}$, which is equal to
($J_{\rm{max}}$+1)($J_{\rm{max}}$+2)/2, increases very rapidly
with the maximum angular momentum $J_{\rm{max}}$. On the other
hand, these same six operators $S^+$ and $Q_M$ are always
 sufficient to obtain all states for arbitrary value of
$J_{\rm{max}}$. The set of building blocks $S^+$ and $Q_M$
allows therefore for a more economic description of the
collective space.

In particular, we may compare different algebraic models by
comparing properties of the corresponding quadrupole operators
$Q$.  In fact, in the algebraic models discussed the monopole
pairs are always equal to the seniority-zero pairs
(cf.{\spc}Eq.{\spc}(\ref{e86a})),
   \begin{equation}\label{e85}
   S^+ = \sqrt{\frac{g}{2}}\frac{1}{\hat{\rr}\hat{\pp}}
         \sum_{jm} a^+_{jm}\tilde{a}^+_{jm} ,
   \end{equation}
with amplitudes independent of $j$. Therefore they are uniquely
determined if the set of $j$-shells is fixed. On the other hand,
in the given space different models have different quadrupole
operators $Q_M$,
   \begin{equation}\label{e87}
   Q_M = h \sum_{j_1j_2} C_{j_1,j_2}
         \left(a^+_{j_1}\tilde{a}_{j_2}\right)^{(2)}_{M} ,
   \end{equation}
defined in terms of structure constants $C_{j_1,j_2}$, see
Eq.{\spc}(\ref{e86b}). Since $j_1$ and $j_2$ can differ by 2 at
most, $Q$ is determined be the set of amplitudes $C_{j,j}$,
$C_{j-1,j}$, and $C_{j-2,j}$.

For $N_0$=6, and for the orthogonal and symplectic models listed
in Table \ref{table1} these amplitudes are presented in
Fig.{\spc}\ref{fig5}. Each of the six algebraic models which can
be defined in this shell-model space is completely determined by
14 amplitudes (or 12 if the $j$=$\frac{1}{2}$ orbital is
excluded) shown in the Figure.  They represent the set of unique
values required by the algebra closure conditions (\ref{eq2.2}).

In principle, the collective space in this shell-model space
should be determined by dynamic conditions starting from
realistic two-body interactions and using an appropriate
effective-operator theory. Such procedure would in general give
quadrupole operators more or less departing from the algebraic
solutions and could either provide these models with a
microscopic justification or disprove.  This programme has not
yet been attempted, probably because of a relative simplicity of
the algebraic models, which allows one to easily obtain
theoretical predictions for experimental quantities, whereas the
difficult question of microscopic justification would require a
much larger effort.

In lack of microscopic derivations of the quadrupole operator,
we compare the algebraic quadrupole operators with the mass
quadrupole-moment operator.  The amplitudes $C_{j_1,j_2}$ of the
$r^2Y_{2M}(\theta,\phi)$ operator reduced to one spherical
harmonic-oscillator shell read
   \begin{equation}\label{e88}
   C_{j_1,j_2} =
   -\frac{\hat{l}_2}{\sqrt{2\pi}} I_{l_1,l_2} (20l_20|l_10)
   (-1)^{l_1+\frac{1}{2}+j_1}
   \frac{\hat{j_1}\hat{j_2}}{\sqrt{2}}
   \left\{\begin{array}{ccc} j_1 & j_2 & 2 \\
                             l_2 & l_1 & \frac{1}{2} \end{array}\right\} ,
   \end{equation}
where $I_{l_1,l_2}$ is the radial integral,
   \begin{equation}\label{e89}
   I_{l_1,l_2} = \int r^2 dr R_{N_0l_1}(r) r^2 R_{N_0l_2}(r)
               = \frac{\hbar}{m\omega}\left\{
       \begin{array}{ll} (N_0+\frac{3}{2}) &
       \quad \mbox{for $|l_1$$-$$l_2|$=0} \\
              -\sqrt{(N_0-\bar{l}+1)(N_0+\bar{l}+2)} &
       \quad \mbox{for $|l_1$$-$$l_2|$=2} \\
       \end{array}\right. ,
   \end{equation}%
and $\bar{l}$=$(l_1$+$l_2)/2$. Since the formula concerns one
oscillator shell, the orbital angular momentum $l$ has a unique
value for every value of $j$. Note that for $l_1$=$l_2$ the
formula gives the root mean squared radii $\langle r^2\rangle$
of the harmonic-oscillator states, independent of $l$. The
harmonic-oscillator amplitudes (\ref{e88}) are plotted in
Fig.{\spc}\ref{fig5} in units of this value of
$\langle{r^2}\rangle$.

For the SD SO(8) and Sp(6) models the diagonal amplitudes
$C_{j,j}$ differ very much from the harmonic oscillator values.
However, when the G, and then I pairs are added, the agreement
becomes better. The same observation also holds for the
amplitudes $C_{j-1,j}$. Here one may in principle change the
sign of every amplitude by an appropriate changes of phases of
single-particle states. For the SDGI models, and also for the
harmonic oscillator, the phases in Eqs.{\spc}(\ref{e86b}) and
(\ref{e88}) are such that these amplitudes are positive. For the
other models one could have fixed the phases in a different way
and change signs of $C_{j-1,j}$, but even then the agreement
with the harmonic oscillator values would not become much
better. For the algebraic models the signs of amplitudes
$C_{j-2,j}$ are different than in the harmonic oscillator.

Of course the harmonic-oscillator amplitudes are not necessarily
the best possible dynamical values for the quadrupole moment
operator in a given major shell. However, this operator has
often been used in the conventional shell model and is known to
provide a fair first approximation for quadrupole collective
phenomena.  Our results indicate that the algebraic models
restricted to S and D pairs are rather remote from the harmonic
oscillator estimates. The desirability to include higher angular
momentum pairs has also been suggested by many other dynamical
analyses, e.g.{\spc}Refs.{\spc}\cite{IT87,DS89}.

\section{Boson realization of the model}
\label{sec4}
\setcounter{equation}{0}

The Dyson boson realization of an arbitrary fermion algebra
composed of pair creation operators, their hermitian conjugates,
and of all independent commutators has been given in
Ref.{\spc}\cite{DGH91}.  For the structure constants of the
so(4${\pp}$+2) and sp(4${\pp}$+2) algebras given in
Eq.{\spc}(\ref{e26}) we obtain the following mapping:
   \begin{eqnalphalabel}{eq3.4}
   F^+_{JM} &\longleftrightarrow& R^{\dagger}_{JM}\equiv gB^{\dagger}_{JM}
      -\frac{2g}{\hat{{\rr}}^2} \sum_{J_1J_2J_3J'}
       \hat{J_1}\hat{J_2}\hat{J_3}\hat{J'}
          \left\{ \begin{array}{ccc} {\pp}   & {\pp}   & J   \\
                                     {\pp}   & {\pp}   & J_3 \\
                                     J_2 & J_1 & J'  \end{array}\right\}
             ((B^{\dagger}_{J_1} B^{\dagger}_{J_2})^{(J')}
                   \tilde{B}_{J_3})^{(J)}_{M} ,
                               \nonumber \\ && \\
                                                        \steplet
   F_{JM} &\longleftrightarrow& B_{JM} , \\
                                                        \steplet
   P_{JM} &\longleftrightarrow& \frac{2h}{\hat{{\rr}}}(-1)^{J+2{\pp}+1}
            \sum_{J_1J_2}\hat{J}_1 \hat{J}_2
            \left\{ \begin{array}{ccc} J_1 & J_2 & J \\
                                        {\pp}  &  {\pp}  & {\pp}
                      \end{array}\right\}
   (B^{\dagger}_{J_1} \tilde{B}_{J_2})^{(J)}_{M} ,
   \end{eqnalphalabel}%
where $\tilde{B}_{JM}$=$(-1)^{J-M}B_{J,-M}$.

The boson mapping of the so(4${\pp}$+2) and sp(4${\pp}$+2)
algebras has the same structure as the generalized Dyson boson
mapping{\spc}\cite{KM91,JDFJ71} of bifermion operators (the notation
is the same as in Ref.{\spc}\cite{GEH86})
   \begin{eqnalphalabel}{eq3.1}
   a^{\alpha} a^{\beta}  &\longleftrightarrow&
     R^{\alpha\beta} = B^{\alpha\beta}
               - B^{\alpha\theta} B^{\beta\rho} B_{\theta\rho} , \\
                                                        \steplet
   a_{\beta}  a_{\alpha} &\longleftrightarrow& B_{\alpha\beta} , \\
                                                        \steplet
   a^{\alpha} a_{\beta}  &\longleftrightarrow&
     B^{\alpha\theta} B_{\beta\theta} .
   \end{eqnalphalabel}%
In this case, each of the single-particle indices
$\alpha,\beta,\ldots$ denotes all quantum numbers of the fermion
creation and annihilation operators, i.e., either
${\rr}m_{\rr}{\pp}m_{\pp}$, or in the vector coupled form,
Eq.{\spc}(\ref{e14a}), ${\rr}{\pp}jm$.  The mapping
(\ref{eq3.1}) can be transformed to the collective variables by
introducing collective amplitudes relevant to our model,
   \begin{equation}\label{eq3.2}
   \chi^{{\RR}{\Pp}J}_{j_1 j_2}=\frac{\sqrt{2}}{\sqrt{1+\delta_{j_1 j_2}}}
   \hat{{\RR}}\hat{{\Pp}}\hat{j}_1\hat{j}_2
   \left\{ \begin{array}{ccc} {\rr}&{\pp}&j_1\\{\rr}&{\pp}&j_2\\{\RR}&{\Pp}&J
   \end{array}\right\} ,
   \end{equation}
which fulfill the orthogonality and completeness relations
   \begin{eqnalphalabel}{eq3.3}
   \sum_{j_1 \leq j_2} \chi^{{\RR}{\Pp}J}_{j_1 j_2}
                       \chi^{{\RR}'{\Pp}'J}_{j_1 j_2} & = &
   \delta_{{\RR}{\RR}'}\delta_{{\Pp}{\Pp}'} , \\
                                                        \steplet
   \sum_{{\RR}{\Pp}} \chi^{{\RR}{\Pp}J}_{j_1 j_2}
                     \chi^{{\RR}{\Pp}J}_{j'_1 j'_2} & = &
   \textstyle{\frac{1}{1+\delta_{j_1 j_2}}}
      \left(\delta_{j_1 j'_1}\delta_{j_2 j'_2}
   + (-1)^{{\RR}+{\Pp}+J+j_1-j_2}\delta_{j_1 j'_2}\delta_{j_2 j'_1}\right) .
   \end{eqnalphalabel}%
The mapping of the spectrum generating algebra ${\cal A}$, which
corresponds to ${\RR}=0$, can be achieved simply by dropping all
other collective amplitudes with ${\RR}\neq 0$ due to the
skeletonization theorem{\spc}\cite{GH81,KV88}, and one obtains
Eqs.{\spc}(\ref{eq3.4}).

Alternatively, we can construct this boson mapping making use of
generalized coherent state  approach{\spc}\cite{Dob81}.  It is based
on a one-to-one correspondence between the coherent states in
the fermion space and in the boson space
   \begin{equation}\label{eq3.6}
   |C\rangle\equiv \exp(\sum C^*_{JM}F^+_{JM})|0\rangle \longleftrightarrow
   |C)\equiv \exp(\sum C^*_{JM}B^{\dagger}_{JM})|0) .
   \end{equation}
The action of boson operators on the coherent state is simply
   \begin{eqnalphalabel}{eq3.7}
   B^{\dagger}_{JM}|C)&=&\frac{\partial}{\partial C^*_{JM}}|C) , \\
                                                        \steplet
   B_{JM}|C)&=&C^*_{JM}|C) .
   \end{eqnalphalabel}%
Similarly the fermion collective pair $F^+_{JM}$ acts as
   \begin{equation}\label{eq3.8}
   F^+_{JM}|C\rangle=\frac{\partial}{\partial C^*_{JM}}|C\rangle .
   \end{equation}
On the other hand the action of the other fermion operators on
the fermion coherent state has to be evaluated using the
commutation relations (\ref{eq2.2}).  Any fermion state
$|\Psi\rangle$ and the corresponding boson state $|\Psi)$ can be
represented using the coherent states as $\langle C|\Psi\rangle$
and $(C|\Psi)$, respectively. Then we can find boson images of
fermion-pair operators identifying
   \begin{equation}\label{eq3.10}
      C_{JM}\longleftrightarrow B^{\dagger}_{JM}\; \;,\;\;\;
   \frac{\partial}{\partial C_{JM}}\longleftrightarrow B_{JM}
   \end{equation}
which results in (\ref{eq3.4}).

\section{Spurious states}
\label{sec6}
\setcounter{equation}{0}

Bosonization of any fermion model from expressions (\ref{eq3.4})
naturally leads to the identification of a corresponding boson
Fock space.  While the so-called physical states, obtained by
repeated action of the images (\ref{eq3.4}a) on the boson
vacuum, still obey restrictions due to the Pauli principle (as
discussed in Sec.{\spc}\ref{sec5}), this does not hold for
general states in the boson space and in particular not for
typical boson basis states of the complete Fock space.
Diagonalization in this basis therefore generally leads to the
occurrence of spurious states.  This situation and the
identification of spurious states is discussed at length in
Refs.{\spc}\cite{GEH86,DGH91,NG93,HJNG93,chungli}

For the SO(12) and Sp(10) models the associated ideal boson
space spans the symmetric representation $[N]$ of the $sdg$
boson U(15) group. In the U(15)$\supset$SU(5) reduction chain,
that is in Sp(10) model, the SU(5) Young labels agree with those
given in the preceding section{\spc}\cite{KJMB87}. In the boson
space, however, there is no Pauli-principle restriction. One
immediately deduces that for $N\leq\frac{\Omega}{5}$ , the
correspondence between fermion states and boson states is
one-to-one. For $N>\frac{\Omega}{5}$ , the most symmetric
bosonic SU(5) representations are unphysical.  In the fermionic
SU(5) space, these representations are forbidden as the most
antisymmetric conjugate representations  are restricted by the
dimensionality of the inert space.

Turning to the SO(12) model, the SU(6) Young labels of the
relevant U(15)$\supset$SU(6) reduction agree with the labels
given in the preceding section for the fermionic active
space{\spc}\cite{KJMB87}. In the boson space, there is again no
Pauli-principle restriction.  For $N>\frac{\Omega}{3}$,
unphysical representations occur in the boson space.

An alternative way to investigate the spurious states discussed
above, is to calculate the overlap of generalized coherent
states (\ref{eq3.6}).  This overlap was calculated for the SO(8)
and the Sp(6) models in Ref.{\spc}\cite{DGH91}.  Tailoring the
discussion in Ref.{\spc}\cite{Dob81} to the coherent state overlap of
the generalized Ginocchio model, one sees that it is possible to
write
\begin{equation}\label{eq5.2}
\langle C,\Omega|C',\Omega\rangle=\left(\langle C,\Omega_0|C',\Omega_0
\rangle\right)^{\Omega/\Omega_0}
\end{equation}
where $\Omega_0=i+\frac{1}{2}$ when $K=0$ and $\Omega_0=2k+1$
when $I=0$. Consequently, one expects spurious states to appear
when $N>\frac{\Omega}{\Omega_0}$, where $N$ is the number of
bosons.

For the SO(12) model analyzed in terms of s-, d-, and g-bosons,
the condition for the appearance of spurious states is thus
$N>\frac{\Omega}{3}$ and for the Sp(10) model{\spc}\cite{FYZ91}
$N>\frac{\Omega}{5}$.  This should be compared (for the s- and
d-boson models) with $N>\frac{\Omega}{2}$ for SO(8)  and
$N>\frac{\Omega}{3}$ for Sp(6).  The condition for the SO(8)
model in fact means that there are no spurious states when
bosons are counted as in IBM, where hole bosons determine the
boson number past mid-shell, implying that the number of bosons
is always less or equal to a quarter of the shell
size{\spc}\cite{Gin80}.  Clearly this one-to-one correspondence
between physical and ideal $sd$-boson states in SO(8) does not
generalize to $sdg$-states in SO(12).

In principle, one could use the U(15)$\supset$SU(5) and
U(15)$\supset$SU(6) classification schemes of the ideal boson
space in the Sp(10) and SO(12) models, respectively, and then
restrict the physical space by throwing out the most symmetric
unphysical SU(5) or SU(6) representations. Algebraic techniques
of the above mentioned classification chains are not, however,
developed in full details. More practical seem to be approaches
in which  d- and g-boson spaces are separated as e.g. in the
U(15)$\supset$U(5)$\otimes$U(9) chain. Then, however, physical
and unphysical components mix in the boson basis states and only
after diagonalization of the Hamiltonian they become separated.

As for the actual identification of spurious states, these may
be recognized by calculating the matrix elements of a general
SO(12) or Sp(10) operator between the eigenstates of whichever
SO(12) or Sp(10) Hamiltonian is being used. This method is based
on the structure{\spc}\cite{GEH86}
\begin{eqnalphalabel}{eq5.4}
({{\tilde\varphi}_{\rm spur}}|\Theta_{\rD}|{\psi_{\rm phys}})&=&0\; ,\\
                                                        \steplet
({{\tilde\psi}_{\rm phys}}|\Theta_{\rD}|{\varphi_{\rm spur}})&\neq&0\; .
\end{eqnalphalabel}%
In other words, the Dyson boson image $\Theta_{\rD}$ of any
fermion operator $\Theta$ does not scatter outside the physical
subspace when acting on a physical ket state.

Another possibility to identify spurious states utilizes the
application of the so-called
${\cal{R}}$-projection{\spc}\cite{Dob81,DGH91}. The
${\cal{R}}$-projector acting on a boson state replaces each
boson creation operator $B^{\dagger}_{JM}$ by the Dyson boson
image $R^{\dagger}_{JM}$ (\ref{eq3.4}). Because spurious
bra-eigenstates are not contaminated by physical components
(which is not true for the spurious ket-eigenstates) a spurious
bra-eigenstate of the boson Hamiltonian is annihilated by the
${\cal{R}}$-projection.  A suitable way to detect spurious states
in this way is to evaluate the norms of the ${\cal{R}}$-projected
bra-eigenstates of the Hamiltonian{\spc}\cite{GNHD92,HJNG93,chungli}.
The norms of spurious states are zero. To calculate the norms
one needs the overlap of ${\cal{R}}$-projected basis states with
non-projected basis states.  This can be computed by iteration
for different boson numbers from the relation
\begin{eqnarray}\label{eq5.5}
{\cal{R}}(N,L;I_a,I_b)&=&\sum_{L' I'_a I'_b}\frac{1}{\hat{L}\hat{L'}}
\frac{1}{n_J} ( N I_b L\| B^{\dagger}_J\|N-1 I'_b L' )
              \nonumber \\ & &
              ( N I_a L\| R^{\dagger}_J\|N-1 I'_a L' )
              {\cal{R}}(N-1,L';I'_a,I'_b)
\end{eqnarray}
Here $R^{\dagger}_J$ is given by Eq.{\spc}(\ref{eq3.4}a) and $J$
should be chosen in such a way that the first reduced matrix
element in (\ref{eq5.5}) is non-zero. The iteration starts from
$N=2$ and ${\cal{R}}(N=1,L;I_a,I_b)=\hat{L}\delta_{ab}$.
To decide whether an eigenstate $|{N,\varphi,L})$
is spurious or not we just calculate
$\sum_{I_b}{\cal{R}}(N,L;I_a,I_b)({N,{\tilde\varphi},L}| N I_b L)$.
This quantity is zero for each $I_a$ if the state is non-physical.

In general, the method (\ref{eq5.4}) is simpler to implement.
One has, however, to calculate quite a number of matrix elements
to identify spurious states unambiguously.  On the other hand,
the ${\cal{R}}$-projection is somewhat more complicated to
implement, but the identification is made from a single matrix
element.  From previous studies{\spc}\cite{GNHD92,HJNG93} we should
point out that the ${\cal{R}}$-projection is more reliable when
one considers a slightly truncated boson system or one with
slightly broken symmetry.

Further application in the SO(12) model can be found in
Ref.{\spc}\cite{chungli} where it is demonstrated that a good fit to a
spectrum from a phenomenological boson analysis does not
necessarily guarantee the absence of large spurious components.

\section{Quadrupole collective spectra in the SD and SDG models}
\label{sec7}
\setcounter{equation}{0}

In this section we present examples of collective spectra which
result from the SDG fermion pair algebraic models and compare
them with the well-known SD solutions.  We consider the $N_0$=6
major shell of normal-parity subshells, where one can construct
the SD Sp(6) and SDG SO(12) models in the
$j$=$\frac{1}{2}$$,\ldots,$$\frac{11}{2}$ space ($\Omega$=21), and
the SD SO(8) and SDG Sp(10) models in the
$j$=$\frac{3}{2}$$,\ldots,$$\frac{11}{2}$ space ($\Omega$=20) (see
Table \ref{table1}).

All the calculations were done after the Dyson boson
mapping had been performed in the ideal $s, d$ or $s, d, g$ space
using modified and generalized interacting boson model codes.
The ideal boson basis was classified according to U(5) chain for $d$-bosons
and U(9) chain for $g$-bosons.

For the SD models we consider the Hamiltonian
   \begin{equation}\label{e80}
   H = (1-x)\left(G_0 S^+S + G_2 D^+\cdot D\right) + x\kappa_2Q\cdot Q ,
   \end{equation}
where $S^+S$ and $D^+$$\cdot$$D$ are the monopole and quadrupole
pairing interactions, respectively, and $Q$$\cdot$$Q$ is the
quadrupole-quadrupole interaction.  In notation of
Sec.{\spc}\ref{sec3}, the pairing and multipole operators are
identified as $S^+$$\equiv$$F^+_{00}$,
$D^+_M$$\equiv$$F^+_{20}$, and $Q_M$$\equiv$$P_{2M}$ with
traditional normalization factors of $g$=$\Omega$ and
$h$=$-$$\sqrt{2\Omega}$.  Parameter $x$ allows for a continuous
change of the Hamiltonian between two limiting cases of a pure
pairing-like interaction ($x$=0) and a pure
quadrupole-quadrupole interaction ($x$=1).

In Figs.{\spc}\ref{fig1} and \ref{fig2} we show the results for
the SO(8) and Sp(6) models, respectively, obtained with the
interaction strengths $G_0$=$-$80~keV, $G_2$=$-$30~keV, and
$\kappa_2$=$-$25~keV. All calculations are  for $N_F$=8
fermions, i.e., for the boson number $N$=4 smaller than
$\frac{\Omega}{2}$ and $\frac{\Omega}{3}$ for the values
$\Omega$=20 or 21 actually used. This means that in these
cases all boson states are physical.  Even if we use the same
strengths parameters in both models, one should note that the
operators in terms of which  the SO(8) and Sp(6) Hamiltonians
are defined are not the same.  The phase spaces are also
different, because the $j$=$\frac{1}{2}$ subshell is excluded in
the SO(8) model for $N_0$=6.  On the other hand, the size of the
fermion space influences the results only through the parameter
$\Omega$ which appears in the boson mapping, and similar spectra
can in fact be obtained for rather different values of $\Omega$.

The left parts of Figs.{\spc}\ref{fig1} and \ref{fig2} show
complete SO(8) and Sp(6) spectra for $x$=0 as functions of the
spin $I$.  In both cases one obtains characteristic vibrational
patterns of quadrupole collective excitations. The ratios of
excitation energies in the yrast bands are close to the
vibrational limit of 2, 3, and 4 for $I$=4, 6 and 8,
respectively (see Table \ref{table2}).  For the SO(8) model, the
Hamiltonian has exact SO(5) dynamical symmetry and many states
of different spins are therefore degenerate. For Sp(6) this is
not the case, but at $x$=0 one can still clearly distinguish
approximate multiplets corresponding to vibrational phonon
excitations.

The centre parts of Figs.{\spc}\ref{fig1} and \ref{fig2} show
the dependence of spectra on the parameter $x$ and exhibit the
characteristic pattern of a phase transition to deformed
structures. The energies of the first $2^+$ excitations decrease
and the ratios of excitations in the yrast bands increase. In
the SO(8) case, the limiting situation of a pure
quadrupole-quadrupole force ($x$=1) corresponds to the SO(6)
dynamical symmetry and to the spectrum of a $\gamma$-unstable
rotor with ratios of 2.5, 4.5, and 7 (Table \ref{table2}). For
Sp(6) the same limit gives the rotational structure resulting
from the SU(3) dynamical symmetry.

We may now compare the results of the SD algebraic models with
those which use the SDG fermion pairs. For the $N_0$=6 major shell
we have calculated the collective spectra in the SO(12) and
Sp(10) models defined in their corresponding phase spaces listed
in Table \ref{table1}.  The boson number $N$=4 is also here not
larger than either $\frac{\Omega}{3}$ or $\frac{\Omega}{5}$ and
therefore all boson states are physical.

It turns out that the results for Hamiltonian (\ref{e80})
exhibit very low lying $4^+$ vibrational states which can be
interpreted as hexadecapole excitations of the system. If one
wants to discuss pure quadrupole structures one may push up the
$L$=4 vibrations by adding an appropriately strong
$G^+$$\cdot$$G$ pairing interaction, where
$G^+_M$$\equiv$$F^+_{4M}$ is the hexadecapole fermion
pair creation operator. Therefore, in what follows we present
results for the Hamiltonian
   \begin{equation}\label{e81}
   H = (1-x)\left(G_0 S^+S + G_2 D^+\cdot D\right)
        + x\kappa_2Q\cdot Q + G_4 G^+\cdot G
   \end{equation}
with coupling strengths $G_0$, $G_2$, and $\kappa_2$ equal to
previous values and $G_4$ being equal to 180~keV and 250~keV for
SO(12) and Sp(10), respectively.  These values of $G_4$ ensure
that the hexadecapole excitations do not appear below 3~MeV.

In Figs.{\spc}\ref{fig3} and \ref{fig4} we show the results for
the SO(12) and Sp(10) models using Hamiltonian (\ref{e81}).  For
$x$=0 we again obtain vibrational patterns with characteristic
multiplets of quadrupole phonon excitations in which all
degeneracies are lifted. In the SO(12) case, we see the
well-isolated lowest hexadecapole state which at $x$=0 starts
slightly above 4~MeV and at $x$=1 goes down to about 3.2~MeV.
When $x$ increases from 0 to 1 one obtains the phase transition
to deformed spectra. The symplectic Sp(10) model now gives
ratios of yrast excitations close to those of a
$\gamma$-unstable rotor (Table \ref{table2}), while the
orthogonal SO(12) model ratios resemble those of a rotor.  This
situation is opposite to what one obtains in the SD models.

\section{Summary and conclusions}
\label{sec8}
\setcounter{equation}{0}

In the present paper we have discussed algebraic collective
models based on the concept of favored pairs of nucleons coupled
to given angular momenta. First of all we have shown that the
Ginocchio construction of such models has its roots in a simple
assumption concerning the angular momentum symmetry.  In fact,
if one wants to replace (in a restricted single-particle space)
the physical angular momentum by another $J$=1 operator, and
keep their matrix elements identical, this space necessarily
splits into a set of $r$-$p$-shells of states for which two
different angular momentum labels are simultaneously valid.
This has the direct consequence of splitting the physical
angular momentum into active and inert parts.

Along these guidelines we have constructed generalized Ginocchio
models for pairs of states with angular momenta varying between
0 and  an arbitrary value of $J_{\rm{max}}$.  This leads to two
families of algebraic models based on the SO(4$p$+2) and
Sp(4$p$+2) symmetries for half-integer and integer $p$,
respectively, where $J_{\rm{max}}$=2[p].  We have also
identified possible subgroup chains and representations as well
as linked the model spaces to the nuclear shell model.

Similarly as in other algebraic models, the structure of basic
building blocks (here the structure of pairs) is fixed by the
algebra closure conditions and takes precedence over dynamical
considerations. We have analyzed this structure by discussing
another set of building blocks, i.e., the monopole pair and the
quadrupole single-particle operator.  The latter has been
compared with the mass quadrupole moment operator acting in one
harmonic oscillator shell. This has demonstrated that the models
based on the S and D fermion pairs lead to quadrupole operators
rather different from the harmonic oscillator estimate. On the
other hand, when the $J$=4, and then the $J$=6 fermion pairs are
included the agreement improves. In this way the algebraic
models may become closer to the traditional shell model
description.

By including higher and higher angular momentum pairs we have to
consider larger and larger algebras, and the usual group theory
techniques used to solve  collective dynamical problems become
more and more cumbersome. In the present paper we have
demonstrated that the boson mapping technique can be a viable
alternative. We have introduced Dyson boson mappings for the
general class of algebraic fermion models considered here and
then diagonalized Hamiltonians for two SDG models based on the
SO(12) and Sp(10) symmetries. In fact the standard computer
codes which have been constructed to solve the sdg-boson IBM
models can be used with minor modifications.

An obvious advantage of this procedure is the fact that one is
not bound to the specific symmetry limits of the algebraic
models and an arbitrary Hamiltonian expressed in terms of
generators (not necessarily Casimir operators) can be rather
easily diagonalized.

As an example, we have applied this technique to Hamiltonians
containing monopole and quadrupole pairing as well as
quadrupole-quadrupole interactions. In particular, the SDG
algebraic models based on the SO(12) and Sp(10) symmetries have
been considered and compared with the well-known SD SO(8) and
Sp(6) models.  By changing the relative interaction strengths
between the pairing and quadrupole-quadrupole terms we have
obtained phase transitions between anharmonic vibrations and
rotations.  Very good rotational structures have been obtained
in the SO(12) model where there is no underlying SU(3) symmetry
limit.  We have found very low-lying hexadecapole vibrations
present in both SDG models, which however could be pushed
up to higher energies by using suitable hexadecapole pairing
interaction.

This work was supported by grants from the Foundation for
Research Development of South Africa, the University of
Stellenbosch, in part by the Polish State Committee for
Scientific Research under Contract No.  20450~91~01, and by
grant No. 202/93/2472 from the Grant Agency of the Czech
Republic.

\section*{Appendix: Algebraic structure of the generalized Gi\-nocchio model}
\label{appA}
\setcounter{equation}{0}
\setcounter{section}{1}
\renewcommand{\theequation}{\Alph{section}.\arabic{equation}}

Fermion pairs of Eq.{\spc}(\ref{e10}), which are the basic building
blocks of the generalized Ginocchio model, can be presented in the
following way:
   \begin{eqnarray}
   F^+_{JM} &=&\sqrt{\frac{g}{2(2\rr+1)}}
         {\displaystyle\sum_{m_{\rr}\geq0}}
         {\displaystyle\frac{1}{1+\delta_{m_{\rr}0}}}
         {\displaystyle\sum_{m_{\pp}m'_{\pp}}}
             ({\pp}m_{\pp}{\pp}m'_{\pp}|JM) \nonumber \\
            &\times& \left(a^+_{{\rr}m_{\rr}{\pp}m_{\pp}}
          \breve{a}^+_{{\rr}m_{\rr}{\pp}m'_{\pp}}
      -(-1)^{2\rr}a^+_{{\rr}m_{\rr}{\pp}m'_{\pp}}
          \breve{a}^+_{{\rr}m_{\rr}{\pp}m_{\pp}}\right) , \label{A1}
   \end{eqnarray}
where
   \begin{equation}\label{A2}
   \breve{a}^+_{{\rr}m_{\rr}{\pp}m_{\pp}}=(-1)^{\rr-m_{\rr}}
           a^+_{{\rr},-m_{\rr}{\pp}m_{\pp}} \quad .
   \end{equation}
For different values of $m_{\rr}$$\geq$0, the terms in
Eq.{\spc}(\ref{A1}) are independent of one another and the
algebra comprised of the pair creation operators $F^+_{JM}$,
their hermitian conjugates and all possible commutators is
therefore a direct sum of algebras for all $m_{\rr}$$\geq$0.

For integer $\rr$ we may further decouple the $a$ and
$\breve{a}$ fermions by using the unitary transformation:
   \begin{equation}\label{A3}
   {a'}^+_{{\rr}m_{\rr}{\pp}m_{\pp}} = \left\{\begin{array}{ll}
      {\displaystyle\frac{1}{\sqrt{2(1+\delta_{m_{\rr}0})}}}
           \left(a^+_{{\rr}m_{\rr}{\pp}m_{\pp}} +
         \breve{a}^+_{{\rr}m_{\rr}{\pp}m_{\pp}}\right)
            &\mbox{for}\quad m_{\rr}\geq0 , \\
      {\displaystyle\frac{i}{\sqrt{2(1+\delta_{m_{\rr}0})}}}
           \left(a^+_{{\rr},-m_{\rr}{\pp}m_{\pp}} -
         \breve{a}^+_{{\rr},-m_{\rr}{\pp}m_{\pp}}\right)
            &\mbox{for}\quad m_{\rr}\leq0 , \end{array}\right.
   \end{equation}
where for $m_{\rr}$=0 one takes the non-zero result.  In terms
of fermions ${a'}^+_{{\rr}m_{\rr}{\pp}m_{\pp}}$, the fermion
pair creation operator is a sum of $2\rr+1$ independent pairs
for different $m_{\rr}$ values, i.e.,
   \begin{equation}\label{A4}
   F^+_{JM} =\sqrt{\frac{g}{2(2\rr+1)}}
         {\displaystyle\sum_{m_{\rr}}}
         {\displaystyle\sum_{m_{\pp}m'_{\pp}}}
             ({\pp}m_{\pp}{\pp}m'_{\pp}|JM)
            {a'}^+_{{\rr}m_{\rr}{\pp}m_{\pp}}
            {a'}^+_{{\rr}m_{\rr}{\pp}m'_{\pp}} .
   \end{equation}
Hence the algebra is the direct sum of $2\rr+1$ identical copies
of the pair algebra for fermions occupying a single-$\pp$ shell.
Since this shell contains $2\pp+1$ states, the algebra
in this case is so(2(2$\pp$+1)).

For half-integer $\rr$ the decoupling of the $a$ and $\breve{a}$
fermions is not possible. However, in this case the terms in
Eq.{\spc}(\ref{A1}) are {\em symmetric} with respect to
exchanging the indices $m_{\pp}$ and $m'_{\pp}$ and therefore
they can be represented as pairs of {\em bosons}. In fact we may
define these boson pairs as
   \begin{equation}\label{A5}
   X^{\dag}_{JM} =\sqrt{\frac{g}{2(2\rr+1)}}
         {\displaystyle\sum_{m_{\rr}>0}} {\ }
         {\displaystyle\sum_{m_{\pp}m'_{\pp}}}
             ({\pp}m_{\pp}{\pp}m'_{\pp}|JM)
              b^{\dag}_{{\rr}m_{\rr}{\pp}m_{\pp}}
              b^{\dag}_{{\rr}m_{\rr}{\pp}m'_{\pp}} ,
   \end{equation}
where $b^{\dag}_{{\rr}m_{\rr}{\pp}m_{\pp}}$ create bosons in the
$(2\rr+1)/2$ copies of the single-$\pp$ shell. Note that in the
case of half-integer $\rr$ there is no $m_{\rr}$=0 term in
Eq.{\spc}(\ref{A5}).

It is now easy to show that the algebra of the fermion pairs
$F^+_{JM}$, $F_{JM}$ and their commutators is identical to the
algebra of the boson pairs $X^{\dag}_{JM}$, $-X_{JM}$, and their
commutators.  It is important to note that one has to change the
sign of the boson pair annihilation operator $X_{JM}$ in order
to obtain the identical algebras. It means that we use different
real forms of both algebras and their representations are of
course not equivalent.  However, the angular momentum coupling
does not depend on which real form is used. Hence the
calculations can be facilitated by working in the
angular-momentum-uncoupled representation and we may use the
commutation relations of boson operators
$b^{\dag}_{\mu}b^{\dag}_{\nu}$ to calculate those of the fermion
operators
$a^+_{\mu}\breve{a}^+_{\nu}+a^+_{\nu}\breve{a}^+_{\mu}$.  For
bosons occupying the $2\pp+1$ states of the single-$\pp$ shell
the algebra of boson-pair operators is  sp(2(2$\pp$+1)) and such
is therefore the generalized Ginocchio algebra for half-integer
$\rr$.

\newcommand{\NP}[1]{{\it Nucl.\ Phys.\ {\bf {#1}}}}
\newcommand{\PR}[1]{{\it Phys.\ Rev.\ {\bf {#1}}}}
\newcommand{\PRC}[1]{{\it Phys.\ Rev. {\bf C{#1}}}}
\newcommand{\PRL}[1]{{\it Phys.\ Rev.\ Lett.\ {\bf {#1}}}}
\newcommand{\PL}[1]{{\it Phys.\ Lett.\ {\bf {#1}}}}
\newcommand{\AP}[1]{{\it Ann.\ Phys.\ (N.Y.) {\bf {#1}}}}
\newcommand{\RMP}[1]{{\it Rev.\ Mod.\ Phys.\ {\bf {#1}}}}
\newcommand{\MF}[1]{{\it Mat.\ Fys.\ Medd.\ Dan.\ Vid.\ Selsk.\ {\bf {#1}}}}
\newcommand{\JMP}[1]{{\it J.\ Math.\ Phys.\ {\bf {#1}}}}
\newcommand{\JPG}[1]{{\it J.\ Phys.\ G: Nucl.\ Part.\ Phys.\ {\bf {#1}}}}

\clearpage

\renewcommand{\arraystretch}{1.3}

\newcommand{  \ih}{\frac {1}{2}}
\newcommand{\iiih}{\frac {3}{2}}
\newcommand{  \vh}{\frac {5}{2}}
\newcommand{\viih}{\frac {7}{2}}
\newcommand{ \ixh}{\frac{ 9}{2}}
\newcommand{ \xih}{\frac{11}{2}}
\newcommand{\rms}{{\rm shell}}
\newcommand{\rma}{{\rm pairs}}
\newcommand{\rmg}{{\rm group}}
\newcommand{\Sp}{{\rm Sp}}
\newcommand{\SO}{{\rm SO}}

\begin{table}\caption[T1]{
The SD, SDG, and SDGI fermion-pair algebraic models in
normal-parity nuclear shells with radial quantum numbers
$N_0$=3,$\ldots,$6, i.e., for particle numbers between 28 and
184. The asterisks denote the cases studied in the FDSM{\spc}\cite{FDSM}.}
\label{table1}
\[\begin{array}{|c|c|c|ccllc|}
\hline
N_0&  \rms  &        j         & {\rr} & {\pp} & \rma & \rmg    & \Omega\\
\hline\cline{1-8}
 3 & 28-50  & \ih  \ldots\vh   & \iiih &   1   & SD   & \Sp(6)^*&    6  \\
   &        &                  &   1   & \iiih & SD   & \SO(8)^*&    6  \\
\hline
 4 & 50-82  & \ih  \ldots\viih &   2   & \iiih & SD   & \SO(8)^*&   10  \\
   &        &                  & \iiih &   2   & SDG  & \Sp(10) &   10  \\
              \cline{3-8}
   &        & \iiih\ldots\viih & \vh   &   1   & SD   & \Sp(6)  &    9  \\
   &        &                  &   1   & \vh   & SDG  & \SO(12) &    9  \\
\hline
 5 &  82-126 & \ih  \ldots\ixh  & \ih;
                                 \viih &   1   & SD   & \Sp(6)^*&   15  \\
   &        &                  & \vh   &   2   & SDG  & \Sp(10) &   15  \\
   &        &                  &   2   & \vh   & SDG  & \SO(12) &   15  \\
              \cline{3-8}
   &        & \iiih\ldots\ixh  &   3   & \iiih & SD   & \SO(8)  &   14  \\
   &        &                  & \iiih &   3   & SDGI & \Sp(14) &   14  \\
\hline
 6 & 126-184 & \ih  \ldots\xih  & \iiih;
                                 \ixh  &   1   & SD   & \Sp(6)^*&   21  \\
   &        &                  &   3   & \vh   & SDG  & \SO(12) &   21  \\
   &        &                  & \vh   &   3   & SDGI & \Sp(14) &   21  \\
              \cline{3-8}
   &        & \iiih\ldots\xih  &  0;
                                    4  & \iiih & SD   & \SO(8)  &   20  \\
   &        &                  & \viih &   2   & SDG  & \Sp(10) &   20  \\
   &        &                  &   2   & \viih & SDGI & \SO(16) &   20  \\
\hline
\end{array}\]
\end{table}

\newcommand{\rmM}{\multicolumn{2}{|@{\hspace{1.0em}}l|@{\hspace{1.0em}}}
                 {{\rm Model}}}
\newcommand{\rmv}{\multicolumn{2}{|@{\hspace{1.0em}}l|@{\hspace{1.0em}}}
                 {{\rm vibrational}}}
\newcommand{\rmr}{\multicolumn{2}{|@{\hspace{1.0em}}l|@{\hspace{1.0em}}}
                 {{\rm rotational}}}
\newcommand{\rmu}{\multicolumn{2}{|@{\hspace{1.0em}}l|@{\hspace{1.0em}}}
                 {\gamma{\rm -unstable}}}

\begin{table}\caption[T2]{
Ratios of excitations
$R_I$=$(E_{I^+}$$-$$E_{0^+})/(E_{2^+}$$-$$E_{0^+})$ in yrast
bands of algebraic SD and SDG models in the vibrational ($x$=0)
and rotational ($x$=1) limit.  Results for the vibrational,
rotational, and $\gamma$-unstable quadrupole model are also
given for comparison.}
\label{table2}
\[\begin{array}{|@{\hspace{1.0em}}ll@{\hspace{1.0em}}|
                   @{\hspace{1.0em}}*{4}{c@{\hspace{1.0em}}}|}
\hline
\rmM              &   I=4  &  I=6  &  I=8  &  I=10   \\
\hline
\rmv              &   2    &  3    &  4    &   5     \\
\hline
\Sp(6)     &  x=0 &   1.91 &  2.74 &  3.69 &         \\
\SO(8)     &  x=0 &   1.89 &  2.68 &  3.26 &         \\
\Sp(10)    &  x=0 &   1.86 &  2.61 &  3.09 &   5.02  \\
\SO(12)    &  x=0 &   1.87 &  2.62 &  3.25 &   5.58  \\
\hline
\rmr              &   3.33 &  7    & 12    &  18.33  \\
\hline
\Sp(6)     &  x=1 &   3.33 &  7    & 12    &         \\
\SO(8)     &  x=1 &   2.5  &  4.5  &  7    &         \\
\Sp(10)    &  x=1 &   2.56 &  4.7  &  7.47 &  11.16  \\
\SO(12)    &  x=1 &   3.16 &  6.47 & 11.19 &  22.40  \\
\hline
\rmu              &   2.5  &  4.5  &  7    &  10     \\
\hline
\end{array}\]
\end{table}

\clearpage

\begin{figure}\caption[F5]{
Amplitudes defining the quadrupole operator $Q$,
Eq.{\spc}(\ref{e87}), in the SD SO(8) and Sp(6), SDG SO(12) and
Sp(10), and SDGI SO(16) and Sp(14) algebraic models constructed
in the $N_0$=6 normal-parity major shell.  The asterisks denote
analogous amplitudes of the mass quadrupole moment in the
$N_0$=6 harmonic-oscillator shell plotted in units of $\langle
r^2\rangle$.}
\label{fig5}
\end{figure}

\begin{figure}\caption[F1]{
Spectra of the SD SO(8) Hamiltonian for $N_F$=8 fermions moving
in the $\Omega$=20 normal-parity major shell corresponding to
the radial quantum number of $N_0$=6. The interaction strengths of
Hamiltonian (\ref{e80}) are $G_0$=$-$80~keV, $G_2$=$-$30~keV,
and $\kappa_2$=$-$25~keV.}
\label{fig1}
\end{figure}

\begin{figure}\caption[F2]{
Same as Fig.{\spc}\ref{fig1} for the SD Sp(6) model and
$\Omega$=21.}
\label{fig2}
\end{figure}

\begin{figure}\caption[F3]{
Spectra of the SDG SO(12) Hamiltonian for $N_F$=8 fermions
moving in the $\Omega$=21 normal-parity major shell
corresponding to the radial quantum number of $N_0$=6. The
interaction strengths of Hamiltonian (\ref{e81}) are
$G_0$=$-$80~keV, $G_2$=$-$30~keV, $G_4$=180~keV, and
$\kappa_2$=$-$25~keV.}
\label{fig3}
\end{figure}

\begin{figure}\caption[F4]{
Same as Fig.{\spc}\ref{fig3} for the SDG Sp(10) model with
$G_4$=250~keV and $\Omega$=20.}
\label{fig4}
\end{figure}


\begin{thebibliography}{99}

\bibitem
          {DGH91} {\sc J. Dobaczewski, H.B. Geyer, and F.J.W. Hahne,}
                  \PRC{44} (1991), 1030.
\bibitem
          {KM91}  {\sc A. Klein and E.R. Marshalek,}
                  \RMP{63} (1991), 375.
\bibitem
          {GEH86} {\sc H.B. Geyer, C.A. Engelbrecht, and F.J.W. Hahne,}
                  \PRC{33} (1986), 1041.
\bibitem
          {JH88} {\sc J. Joubert and F.J.W. Hahne,}
                 {\it S. Afr. J. Phys.} {\bf 11} (1988), 51.
\bibitem
          {TT86} {\sc K. Takada and S. Tazaki,}
                 \NP{A448} (1986), 56.
\bibitem
          {TY89} {\sc K. Takada and K. Yamada,}
                 \NP{A496} (1989), 224.
\bibitem
          {TS91} {\sc K. Takada and R. Shimuzu,}
                 \NP{A523} (1991), 354.
\bibitem
          {GNHD92} {\sc H.B. Geyer, P. Navr\'atil, F.J.W. Hahne, and
                   J. Dobaczewski,} {\it in}
                   ``Proceedings of the 4th International Spring
                   Seminar on Nuclear Physics" (A. Covello, Ed.),
                   p.\ 281,  World Scientific, Singapore, 1992.
\bibitem
          {HJNG93} {\sc F.J.W. Hahne, J. Joubert, P. Navr\'atil, and
                   H.B. Geyer,}
                   ``Boson analyses in the Ge isotopes,"
                   {\it Phys. Rev. C} (submitted).
\bibitem
          {FDSM}
                 {\sc C.-L. Wu, D.H. Feng, X.-G. Chen, J.-Q. Chen,
                 and M.W. Guidry,}
                 \PL{168B} (1986), 313;
                 \PRC{36} (1987), 1157;
                 {\sc C.-L. Wu, D.H. Feng, and M.W. Guidry,} to
                 be published in {\it Adv. in Nucl. Phys.}
\bibitem
          {FYZ91}  {\sc P. Feng, C. Yufang, and P. Zhenyong,}
                   {\it Chinese Phys. Lett.} {\bf 8} (1991), 5.
\bibitem
          {DNG93} {\sc J. Dobe\v{s}, P.  Navr\'atil, and H.B.  Geyer,}
                  ``Boson mappings of the fermion
                  dynamical symmetry model,"
                  {\it Phys. Rev. C} (submitted).
\bibitem
          {Dob87} {\sc J. Dobaczewski,}
                  {\it Ann. Univ. M.~Curie-Sk{\l}odowska},
                  Lublin, {\bf XL/XLI, 9} Sect. AAA (1986), 81.
\bibitem
          {NG93}  {\sc P. Navr\'atil and  H.B. Geyer,}
                  \NP{A556} (1993), 165.
\bibitem
          {SGA88} {\sc A. Bohm, Y. Ne'eman, A.O. Barut, and others,}
                  ``Dynamical Groups and Spectrum Generating Algebras",
                  World Scientific, Singapore, 1988.
\bibitem
          {Dob90} {\sc J.  Dobaczewski,}
                  \NP{A506} (1990), 293.
\bibitem
          {VMK88} {\sc D.A Varshalovitch, A.N. Moskalev,
                  and V.K. Kersonskii,}
                  ``Quantum Theory of Angular Momentum",
                  World Scientific, Singapore, 1988.
\bibitem
          {WBC73} {\sc L. Weaver, L.C. Biedenharn, and R.Y. Cusson,}
                  \AP{77} (1973), 250.
\bibitem
          {RR77}  {\sc G. Rosensteel and D.J. Rowe,}
                  \PRL{38} (1977), 10;
                  \AP{126} (1980), 343.
\bibitem
          {Gin80} {\sc J.N. Ginocchio,}
                  \AP{126} (1980), 234.
\bibitem
          {HA69}  {\sc K.T. Hecht and A. Adler,}
                  \NP{A137} (1969), 129.
\bibitem
          {CCK93} {\sc J.Q. Chen, B.Q. Chen, and A. Klein,}
                  \NP{A554} (1993), 61.
\bibitem
          {Dob90a} {\sc J. Dobaczewski,}
                   \PL{241B} (1990), 289.
\bibitem
          {DR90} {\sc J. Dobaczewski and S.G. Rohozi\'nski,} {\it in}
                 ``Proceedings of the 3rd International Spring
                 Seminar on Nuclear Physics", (A. Covello, Ed.),
                 p.\ 351, World Scientific, Singapore, 1990.
\bibitem
          {IT87} {\sc F. Iachello and I. Talmi,}
                 \RMP{59} (1987), 339,
                 and references therein.
\bibitem
          {DS89} {\sc J. Dobaczewski and J. Skalski,}
                 \PRC{40} (1989), 1025.
\bibitem
          {JDFJ71} {\sc D. Janssen, F. D\"onau, S. Frauendorf,
                   and R.V. Jolos,}
                   \NP{A172} (1971), 145;
                   {\sc P. Ring and P. Schuck,}
                   \PRC{16} (1977), 801.
\bibitem
          {GH81}  {\sc H.B. Geyer and F.J.W. Hahne,}
                  \NP{A363} (1981), 45.
\bibitem
          {KV88}  {\sc G.-K. Kim and C.M. Vincent,}
                  \PRC{37} (1988), 2176.
\bibitem
          {Dob81} {\sc J. Dobaczewski,}
                  \NP{A369} (1981), 213, 237;
                  \NP{A380} (1982), 1.
\bibitem
          {chungli} {\sc P. Navr\'atil, H.B. Geyer, J. Dobe\v{s}, and
                    J. Dobaczewski,} {\it in}
                    ``Proceedings of the International Symposium
                    on Nuclear Structure Physics Today", Chung Li,
                    Taiwan,
                    {\it Nucl. Phys. A} (1994), in press.
\bibitem
          {KJMB87} {\sc V.K.B. Kota, J.Van der Jeugt, H.De Meyer, and
                   G.Vanden Berghe,}
                   \JMP{28} (1987), 1644.
\end{thebibliography}
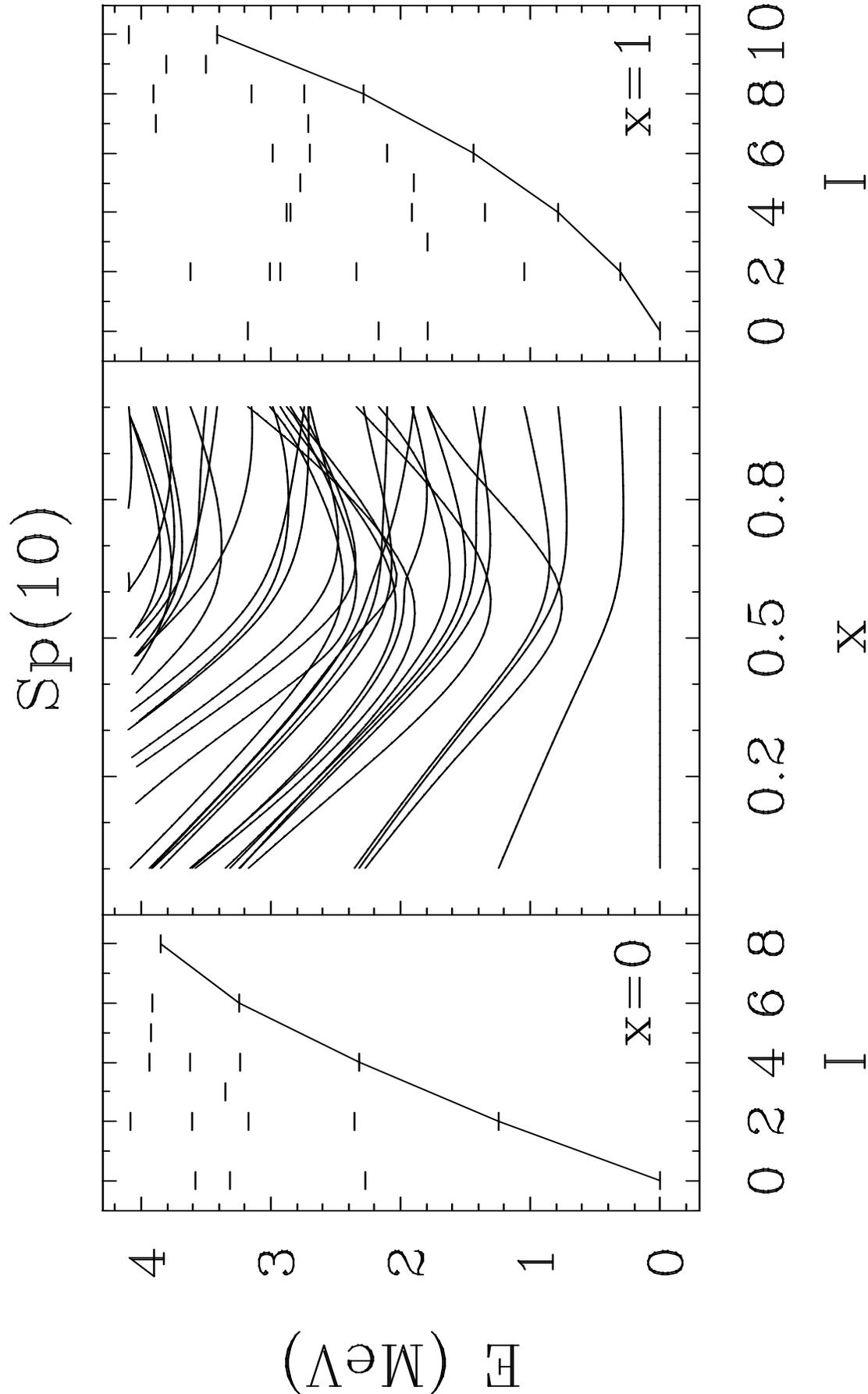
\end{document}